\documentclass[journal]{IEEEtran}

\ifCLASSINFOpdf
\else
   \usepackage[dvips]{graphicx}
\fi
\usepackage{url}

\usepackage{graphicx}
\usepackage{hyperref}
\usepackage{amsmath}
\usepackage{multirow}
\usepackage{booktabs}
\usepackage{amssymb}
\usepackage{color}
\usepackage{orcidlink}

\begin{document}

\title{Neural Scoring: A Refreshed End-to-End Approach for Speaker Recognition in Complex Conditions}

\author{Wan Lin\textsuperscript{\orcidlink{0009-0004-5923-5979}}, Junhui Chen\textsuperscript{\orcidlink{0009-0006-9097-0167}}, Tianhao Wang\textsuperscript{\orcidlink{0009-0004-9915-1235}}, Zhenyu Zhou\textsuperscript{\orcidlink{0009-0009-8492-5222}}, Lantian Li\textsuperscript{\orcidlink{0000-0002-5546-8060}}, Dong Wang\textsuperscript{\orcidlink{0000-0002-1286-0644}}
\thanks{W.L. and D.W. are with the Center for Speech and Language
Technologies (CSLT), BNRist at Tsinghua University, Beijing, China (E-mail: liwan@cslt.org, wangdong99@mails.tsinghua.edu.cn).}
\thanks{J.C., T.W., Z.Z. and L.L. are with the School of Artificial Intelligence, Beijing University of Posts and Telecommunications, Beijing, China (E-mail: chenjh@cslt.org, wangth@bupt.edu.cn, zhouzy@cslt.org, lilt@bupt.edu.cn).}}

\markboth{Journal of \LaTeX\ Class Files, Vol. 14, No. 8, July 2025}
{Shell \MakeLowercase{\textit{et al.}}: Bare Demo of IEEEtran.cls for IEEE Journals}
\maketitle

\begin{abstract}
Modern speaker verification systems primarily rely on speaker embeddings, followed by verification based on cosine similarity between the embedding vectors of the enrollment and test utterances.
While effective, these methods struggle with multi-talker speech due to the \emph{unidentifiability} of embedding vectors.
In this paper, we propose \emph{Neural Scoring} (NS), a refreshed end-to-end framework that directly estimates verification posterior probabilities without relying on test-side embeddings, making it more robust to complex conditions, e.g., with multiple talkers.
To make the training of such an end-to-end model more efficient, we introduce a large-scale trial e2e training (LtE2E) strategy, where each test utterance pairs with a set of enrolled speakers, thus enabling the processing of large-scale verification trials per batch.
Experiments on the VoxCeleb dataset demonstrate that NS consistently outperforms both the baseline and competitive methods across various conditions, achieving an overall 70.36\% reduction in EER compared to the baseline.
\end{abstract}

\begin{IEEEkeywords}
end-to-end model, neural scoring, speaker verification
\end{IEEEkeywords}

\IEEEpeerreviewmaketitle

\section{Introduction}

\IEEEPARstart{S}{peaker} verification (SV) aims to determine whether a given test utterance belongs to a previously enrolled speaker. 
Contemporary SV systems are predominantly built upon the speaker embedding paradigm, exemplified by the x-vector framework~\cite{snyder2018x} and its variants~\cite{zeinali2019but,villalba2020state,li2022cn,chen2022build,qin2022simple}. 
In this paradigm, a variable-length utterance is encoded by a deep neural network (DNN) into a fixed-dimensional vector, termed a speaker embedding. 
Speaker similarity is then quantified by computing the cosine distance between the embeddings of two utterances~\cite{dehak2009support}.
Extensive research has advanced this framework in terms of embedding architectures~\cite{wang2023cam++,wang2023lightweight,alam2023hybrid,liu2024masv,heo2024next}, objective functions~\cite{li2022adaptive,han2023exploring,liu2023awlloss,cai2024leveraging}, and training heuristics~\cite{zhang2023adaptive,cai2023pretraining,choi2025trainable}.

Despite its success, the embedding-based approach faces inherent limitations in scenarios involving multiple speakers—whether overlapping or concatenated speech. In such conditions, the extracted embedding becomes \emph{unidentifiable}: it no longer reliably corresponds to any single speaker in the mixture. Instead, it is often biased by factors such as relative
energy levels and speaker durations, prioritizing dominant acoustic features.
We elaborate on this \textit{embedding unidentifiability} issue in Section~\ref{sec:method}.

Several methods have been proposed to mitigate this issue. 
Most seek to extract the target speech using the enrolled speaker embedding as a bias~\cite{snyder2019speaker,Rao2019TargetSE,zhang2021towards}, or attempt to derive target-aware embeddings~\cite{Shi2021,zhang2022enroll}. 
Although these methods offer partial solutions, they remain constrained by functional limitations of the embedding paradigm, such as embedding indistinguishability under dense overlaps and non-generalizable scoring mechanism across diverse environments.

\begin{figure*}[!htbp]
  \centering
  \vspace{-4mm}
  \includegraphics[width=1\linewidth]{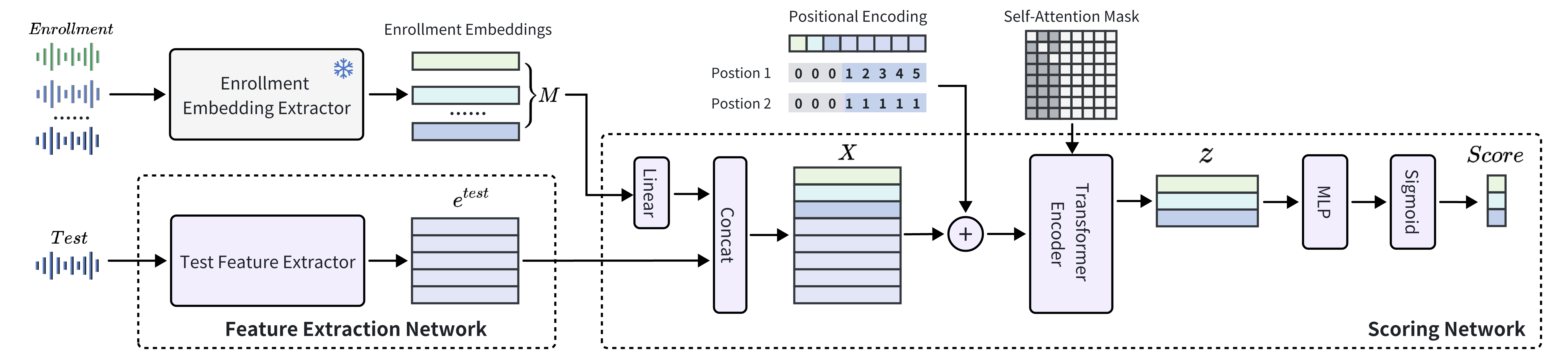}
  \vspace{-6mm}
  \caption{\textit{Overview of the Neural Scoring framework.}}
  \label{fig:structure}
  \vspace{-2mm}
\end{figure*}

To overcome these limitations, we propose \emph{Neural Scoring} (NS), a refreshed end-to-end framework for robust speaker verification under complex conditions.
As illustrated in Fig.~\ref{fig:structure}: it first extracts frame-level features from the test utterance and then estimates the posterior probability that a particular enrolled speaker is present, using a Transformer-based scoring network.
Unlike the embedding paradigm, NS formulates verification as a discriminative scoring problem, enabling it to learn a flexible, task-specific decision boundary and thus circumvent the unidentifiability issue of intermediate embeddings.
This architecture is inherently more general than the conventional embedding paradigm, assuming the neural components are sufficiently expressive and adequately trained.

Although the idea of end-to-end SV dates back to Heigold et al.~\cite{heigold2016end}, it has not gained widespread adoption due to the inherent difficulty of training such models~\cite{wen2022sphereface2}. 
To address this, we introduce a large-scale trial e2e training (LtE2E) strategy, where each test utterance is paired with a set of enrolled speakers, enabling the model to process a large number of verification trials per batch. 
This strategy provides stronger and denser supervision signals, allowing the NS model to achieve stable convergence and better learn the decision boundary under acoustically challenging conditions. 
As we will demonstrate, this training scheme is essential to making the NS framework competitive and robust.

\section{Related Work}

To address multi-talker speech, Thienpondt et al.~\cite{thienpondt2023margin} employed the Mixup technique~\cite{zhang2017mixup}, training the SV model on mixtures of utterances with mixed speaker labels.
While this improves robustness by exposing the model to multi-speaker signals, it does not resolve the fundamental unidentifiability issue inherent in embeddings.

A more direct approach is to incorporate target speaker information during test-time embedding extraction. For example, Snyder et al.~\cite{snyder2019speaker} used speaker diarization to isolate segments corresponding to the enrolled speaker—a strategy also adopted in~\cite{villalba2019state}. While effective in scenarios with concatenated or sequential speakers, diarization often fails in overlapping speech, where speaker boundaries are not cleanly separable.

Other approaches rely on target-aware processing. Zhang et al.~\cite{zhang2021towards} proposed TASE-SVNet, which applies speaker-conditioned enhancement prior to verification. A similar enhancement-based strategy was used in~\cite{Rao2019TargetSE}, though at the cost of increased model complexity. 
Zhang et al.~\cite{zhang2022enroll} introduced EA-ASP, which integrates the enrollment embedding into the pooling layer to emphasize target speaker characteristics in the extracted representation.

Despite methodological differences, these approaches remain fundamentally tied to the embedding-based paradigm, relying on intermediate speaker representations and predefined scoring back-ends. 
A rare exception is Aloradi et al.~\cite{aloradi2022speaker}, who formulated SV as a binary classification task by injecting the enrollment embedding into the feature extractor, thus bypassing the need for a separate scoring back-end. However, the reported performance lags far behind state-of-the-art (SOTA) systems, highlighting the need for further research into the design and training of effective end-to-end SV models.

\section{Neural Scoring}
\label{sec:method}

In this section, we examine the unidentifiability issue associated with the embedding-based approach and introduce the proposed Neural Scoring (NS) framework.

\subsection{Unidentifiability in Embedding-Based Systems}
\vspace{-3mm}
\begin{figure}[!htbp]
  \centering
  \includegraphics[width=0.6\linewidth]{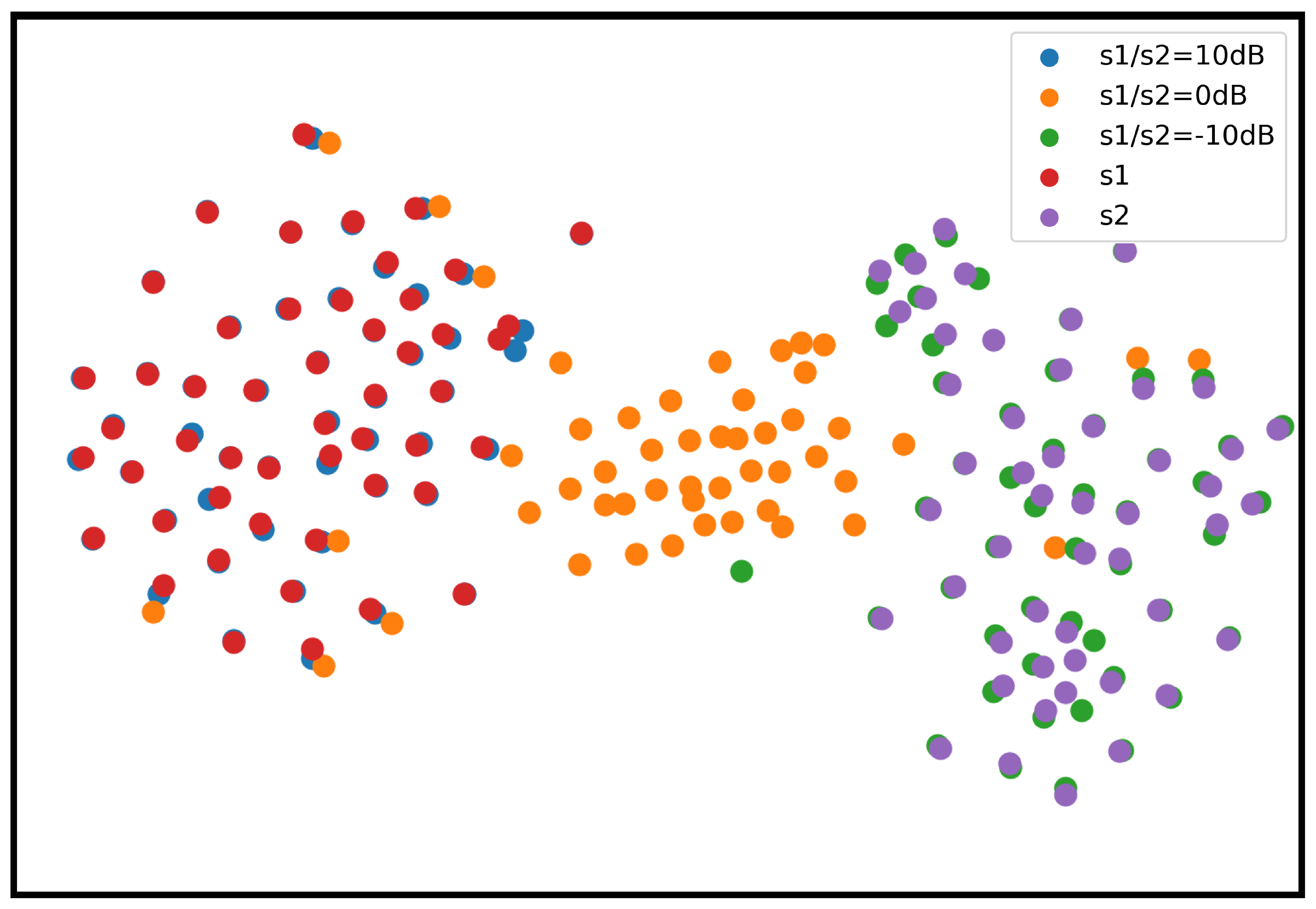}
  \vspace{-2mm}
  \caption{\textit{T-SNE~\cite{van2008visualizing} visualization of speaker embeddings for mixed speech. Two utterances from different speakers in VoxCeleb1~\cite{nagrani2017voxceleb} are mixed at varying SNRs with embeddings extracted using a pre-trained r-vector model~\cite{zeinali2019but}.}}
  \label{fig:tsne}
\end{figure}
Fig.~\ref{fig:tsne} illustrates the behavior of speaker embeddings when two speakers are mixed at different energy ratios. As shown, the resulting embeddings tend to reflect the dominant speaker, i.e., the one with higher energy. When the two speakers contribute equally (i.e., similar energy levels), the embedding deviates from both speakers' true subspaces, appearing as if it belongs to an entirely different identity.

This phenomenon demonstrates the \emph{unidentifiability} problem: embedding representations are ill-defined for multi-talker speech, as they do not faithfully represent either speaker. Similar issues arise when the test utterance is corrupted by human-like noise (e.g., babble), which introduces conflicting speaker cues and leads to degraded embedding quality.

\subsection{The Architecture of NS}

To address this issue, we propose NS, a refreshed end-to-end speaker verification architecture that avoids reliance on embedding similarity. As shown in Fig.~\ref{fig:structure}, the NS framework follows the enrollment-verification SV process and consists of two primary components: an \textbf{enrollment embedding extractor} and a \textbf{verification scoring parser}, which includes a feature extraction network and a scoring network.

\noindent \textit{(1) Enrollment embedding extractor}

This module extracts the enrollment embedding to represent the target speaker, serving as input to the scoring network. 
We adopt a frozen pre-trained r-vector~\cite{zeinali2019but} model as the embedding extractor, ensuring that the enrollment embedding remains consistent and speaker-discriminative.

\noindent \textit{(2) Verification scoring parser}

\textit{(a) Feature extraction network}:  
This module processes the test utterance and produces frame-level features. 
We employ a ResNet34 backbone that is identical to the enrollment embedding extractor. After obtaining the $T$-frame features $e^{test} \in \mathbb{R}^{T \times F \times C}$, we first flatten the frequency $(F)$ and channel $(C)$ dimensions, and then linearly projected them into a $D$-dimensional space, yielding $e^{test} \in \mathbb{R}^{T \times D}$.

\textit{(b) Scoring network}:  
This module estimates the posterior probability that a claimed speaker is present in the test utterance, using enrollment embedding and frame-level features.
We adopt a Transformer~\cite{vaswani2017attention} structure for this process.

Firstly, the enrollment embedding is projected into $D$-dim space by a linear transformation and then concatenated with the frame features of the test utterance. 
This produces a vector sequence $X \in \mathbb{R}^{(1+T) \times D}$, where the first element is the projected enrollment embedding. After adding positional encoding, the resulting sequence is fed into a Transformer encoder. From its output, the latent vector $z$, corresponding to the enrollment embedding (position 0), is extracted and then processed through an MLP (three linear layers with ReLU activations) followed by a Sigmoid function, generating the verification score.
This vector aggregates information across all test frames conditioned on the enrollment embedding, effectively modeling the verification decision.

Moreover, to incorporate positional information, a two-dimensional positional encoding scheme is used. 
The first dimension encodes the absolute position of test frames (with the enrollment embedding assigned position 0) using sinusoidal encoding~\cite{vaswani2017attention}, while the second dimension distinguishes between enrollment embeddings and test frames via learnable binary embeddings (0 for enrollment, 1 for test frames).


\vspace{-2mm}
\subsection{Large-scale Trial E2E Training (LtE2E)}

We illustrate the LtE2E strategy with a batch example. 
Each batch processes $N$ test utterances, where each is paired with a set of $M$ enrollment utterances, resulting in $N \times M$ large-scale verification trials. 
Among $M$ enrollments for a given test utterance, $m$ correspond to target speakers (i.e., speakers present in the test utterance), while the remaining $M-m$ belong to non-target speakers.
To simplify, only $mN$ enrollments are loaded per batch, with non-target trials constructed by reusing target enrollments from other test utterances.

With this setup, the frame features of each test utterance are concatenated with $M$ projected enrollment embeddings in the scoring network, forming $X \in \mathbb{R}^{(M+T) \times D}$. A masking matrix is applied to the self-attention layers to ensure independence between enrollment branches. This design not only reduces training costs, but also enables efficient one-to-many verification through a single forward pass during inference.

To optimize the NS model over all $N \times M$ trials, a weighted binary cross-entropy (BCE) loss is applied:
\begin{equation}
\small
\mathcal{L} = -\frac{1}{NM}\sum_{i,j} \left[ \lambda y_j^i \log r_j^i + (1-\lambda)(1-y_j^i) \log(1-r_j^i) \right],
\end{equation}
\noindent where $r_j^i$ denotes the predicted score for the $i$-th test utterance and $j$-th enrollment, $y_j^i$ indicates whether they match, and $\lambda$ balances the contributions of target and non-target trials.



%

\begin{table*}[!htbp]
\centering
\vspace{-4mm}
\caption{Performance on Vox1-E under different test conditions. The Overall condition includes all trials from the other five conditions. Note that under the single-scenario setting, some test configurations are not applicable to certain models.}
\vspace{-2mm}
\scalebox{0.82}{
\begin{tabular}{lcccccccccccccc}
\toprule
    &       &      & \multicolumn{2}{c}{Clean}      & \multicolumn{2}{c}{Noisy}            & \multicolumn{2}{c}{Concatenation}           & \multicolumn{2}{c}{Overlap}          & \multicolumn{2}{c}{Mixing}              & \multicolumn{2}{c}{Overall} \\ 
\cmidrule(r){4-5} \cmidrule(r){6-7} \cmidrule(r){8-9} \cmidrule(r){10-11} \cmidrule(r){12-13} \cmidrule(r){14-15}
\multirow{-2}{*}{Model} & \multirow{-2}{*}{Params} & \multirow{-2}{*}{FLOPs} & EER(\%)        & minDCF         & EER(\%)              & minDCF               & EER(\%)              & minDCF               & EER(\%)              & minDCF               & EER(\%)              & minDCF               & EER(\%)          & minDCF          \\ 
\midrule
Baseline    & 8.0M    & 9.23G & \textbf{1.11} & \textbf{0.110} & 3.83                & 0.261       & 7.60                & 0.543                & 12.62               & 0.605                & 14.62               & 0.664                & 8.67            & 0.442           \\ 
\midrule
\multicolumn{3}{l}{}    & \multicolumn{12}{c}{Multi-Scenario Training (Main Results)}  \\ 
\midrule
Baseline (random)       & 8.0M      & 9.23G     & 1.47          & 0.144          & 3.55          & 0.310          & 4.30            & 0.436           & 6.32          & 0.520          & 7.84          & 0.600          & 5.00          & 0.414          \\
Margin-Mixup            & 8.0M      & 9.23G     & 1.75          & 0.172          & 4.10          & 0.351          & 5.67            & 0.510           & 7.28          & 0.574          & 7.54          & 0.608          & 5.87          & 0.466          \\
Diarization             & 8.0M      & 13.76G    & 2.11          & 0.213          & 4.74          & 0.390          & 2.63            & 0.229           & 6.63          & 0.459          & 8.02          & 0.610          & 5.83          & 0.412          \\
EA-ASP                  & 5.9M      & 9.24G     & 2.36          & 0.216          & 4.21          & 0.319          & 3.51            & 0.289           & 5.63          & 0.408          & 8.07          & 0.546          & 4.89          & 0.370          \\
\textbf{Neural Scoring} & 6.7M      & 9.29G     & \textbf{1.26}          & \textbf{0.143}          & \textbf{2.28}          & \textbf{0.261}          & \textbf{2.01}            & \textbf{0.206}           & \textbf{3.04} & \textbf{0.293} & \textbf{4.16} & \textbf{0.390} & \textbf{2.57} & \textbf{0.263} \\
\midrule
\multicolumn{3}{l}{}    & \multicolumn{12}{c}{Single-Scenario Training} \\ 
\midrule
Baseline (random)        & -              & -              & -              & -              & 3.82          & 0.262          & 4.16            & 0.480           & 7.06          & 0.628          & 8.20          & 0.646          & -             & -              \\
Margin-Mixup            & -              & -              & -              & -              & -           & -          & 4.42            & 0.435           & 6.64          & 0.531          & 7.77          & 0.586          & -             & -              \\
Diarization             & -              & -              & -          & -          & 3.64          & 0.296          & 2.06            & \textbf{0.186}  & 5.88          & 0.404          & 7.91          & 0.567          & -             & -              \\
EA-ASP                  & -              & -              & 1.81          & 0.191          & 2.93          & 0.310          & 5.31            & 0.519           & 4.97          & 0.382          & 6.84          & 0.511          & -             & -              \\
\textbf{Neural Scoring}         & -              & -              & 1.31          & 0.153          & \textbf{2.20} & \textbf{0.258} & \textbf{1.98}   & 0.214           & \textbf{3.07}          & \textbf{0.302}          & \textbf{4.20}          & \textbf{0.411}          & -             & -               \\ 
\bottomrule
\end{tabular}}
\label{tab:multi}
\vspace{-4mm}
\end{table*}

\vspace{-1.5mm}
\section{Experimental Setup}

\vspace{-1mm}
\subsection{Datasets}

We conducted experiments using VoxCeleb2~\cite{chung2018voxceleb2} for training and VoxCeleb1~\cite{nagrani2017voxceleb} for evaluation. In addition to the original (clean) data, we created four corrupted variants, each constructed by mixing the target speech with a single interference signal at SNRs sampled from $[-3,3]$ dB.

\begin{itemize}

\item \textbf{\textit{Noisy}}: Noise segments were sampled from the MUSAN dataset~\cite{snyder2015musan} and added to the original speech.

\item \textbf{\textit{Concatenation}}: Speech from target and non-target speakers were concatenated sequentially, creating multi-speaker utterances with clear speaker boundaries.

\item \textbf{\textit{Overlap}}: Speech from target and non-target speakers was partially overlapped. The overlap ratio—defined as the proportion of overlapping duration to total utterance length—was randomly sampled from $[0.1, 0.9]$.

\item \textbf{\textit{Mixing}}: Speech from target and non-target speakers were fully overlapped from the start. The shorter utterance was repeated until it matched the length of the longer one.

\end{itemize}

\vspace{-2mm}
\subsection{Settings}

Six models, including our proposed NS framework, were constructed for comparative evaluation. 
All models employed 80-dim Fbanks and produced 256-dim speaker embeddings if applicable. 
The batch size $N$ was set to 256, and final model parameters were obtained by averaging the weights from the last 10 training epochs. Details of each system are as follows:

\begin{itemize}

\item \textbf{\textit{Baseline}}: A standard r-vector model following~\cite{zeinali2019but}, trained solely on the original VoxCeleb2 dataset. 

\item \textbf{\textit{Baseline (random)}}: The same structure as the Baseline system, but trained on both original and corrupted data. For multi-speaker utterances, one speaker was randomly assigned as the target label, reflecting weak supervision.

\item \textbf{\textit{Margin-Mixup}~\cite{thienpondt2023margin}}: The same structure as the Baseline system, but applies the Mixup technique~\cite{zhang2017mixup} with a margin-based regularization. The energy ratio of the two speakers was used as the mixing weight.

\item \textbf{\textit{Diarization}~\cite{snyder2019speaker}}: Extends the Margin-Mixup system by incorporating speaker diarization. Test utterances are segmented into overlapping 1.5-s windows (0.75-s shift). The extracted segment embeddings are clustered into two clusters using Agglomerative Hierarchical Clustering (AHC)~\cite{murtagh2012algorithms}. The mean embedding with higher cosine similarity to the enrollment serves as the test embedding.

\item \textbf{\textit{EA-ASP}~\cite{zhang2022enroll}}: The same structure as the Baseline system, but the enrollment embedding was forwarded to the pooling layer to focus on target-relevant frames and channels.

\item \textbf{\textit{Neural Scoring (NS)}}: Our proposed model. The baseline model was used to produce enrollment embeddings and initialize the test feature extractor. The scoring network consists of a single Transformer layer with 4 attention heads, a hidden size of 256, and a feedforward dimension of 512. During training, $M$ was set to 200, $m$ to 1 for single-scenario training in clean and noisy conditions, and 2 for all others. The $\lambda$ parameter in $\mathcal{L}$ was set to 0.95.

\end{itemize}

\vspace{-1mm}
\section{Results}

Table~\ref{tab:multi} reports the results under two training settings:
(1) \textbf{Multi-Scenario Training}, which combines the original data with four corrupted versions; and
(2) \textbf{Single-Scenario Training}, where models are trained and tested on the same individual scenario.
System performance is evaluated on EER and minimum Detection Cost Function (minDCF) with $P_{tar}$ = 0.01 and $C_{miss}$ = $C_{fa}$ = 1.
Due to space constraints, we report results only on VoxCeleb-E. The complete results, evaluation sets, and code are available at \href{https://github.com/asip-cslt/NS-SV}{NS-SV.git}.

\vspace{-2mm}
\subsection{Multi-Scenario Training}

We begin by examining the results under multi-scenario training, which represents the primary evaluation setting aimed at simulating real-world deployment.

The proposed NS model achieves the best performance across all test conditions, with the only exception being the clean scenario, where the Baseline performs slightly better. In the overall evaluation, NS yields an EER of 2.57\% and a minDCF of 0.263, corresponding to: (1) a 70.36\% relative reduction in EER compared to the Baseline, (2) a 48.60\% reduction compared to Baseline (random), and (3) clear improvements over all other competitive approaches.

In particular, NS exhibits exceptional robustness under corrupted scenarios:
(1) For noisy speech, NS achieves an EER of 2.28\%, showcasing superior performance;
(2) For concatenation, 2.01\% EER, the best among all models;
(3) For overlapping speech, 3.04\% EER, far outperforming other models (e.g., EA-ASP: 5.63\%, Margin-Mixup: 7.28\%);
(4) For the most challenging mixing condition, NS nearly halves the EER compared to Baseline (random) (4.16\% vs. 7.84\%).

Although the Baseline achieves a slightly better result in the clean test condition (1.11\% vs. 1.26\%), this minor trade-off is acceptable given NS’s substantial gains in more challenging and realistic scenarios. 
Notably, NS achieves these improvements with fewer parameters (6.7M vs. 8M) and comparable FLOPs (9.29G vs. 9.23G), demonstrating its efficiency alongside robust performance. 
While EA-ASP uses the fewest parameters among all models (5.9M), its overall performance falls significantly short of NS (e.g., Overall 4.89\% vs. 2.57\%).


\vspace{-2mm}
\subsection{Single-Scenario Training}

Under the single-scenario training setup, the NS model again demonstrates superior performance, maintaining its lead across nearly all test conditions. 
The only exception occurs in the concatenation condition, where the Diarization model achieves a marginally better minDCF (0.186 vs. 0.214), which can be attributed to its design focus on speaker separation in conversations. 
However, the computational cost of the Diarization model is substantially higher than that of all other models, requiring 13.76G FLOPs.


More importantly, cross-comparison between the multi- and single-scenario settings reveals an important insight: NS benefits significantly from multi-scenario training, whereas most other models experience performance degradation. 
For example, the EA-ASP model’s performance drops under multi-scenario training (e.g., Mixing EER: 6.84\% vs. 8.07\%), while NS improves across the board. This indicates that NS is better suited for learning from diverse and complex environments, making it more robust for real-world applications.

\vspace{-2mm}
\subsection{Ablation Study}

Table~\ref{tab:ablation} presents the results of ablation studies evaluating the impact of each architectural component as well as the effectiveness of the LtE2E training strategy.

\begin{itemize}
  \item \textbf{With shared encoder}: Evaluates performance when enrollment and test share the same feature extractor. It can be observed that using a shared encoder consistently reduces performance, demonstrating the importance of enrollment-test decoupling. 
  
  \item \textbf{With multiple layers}: Evaluates performance when using 8 Transformer encoder layers. A clear performance drop is observed, indicating that a single layer is sufficient to model the cross-relation between enrollment embeddings and test frame features. Excessive network complexity may lead to overfitting and reduced generalizability.
  
  \item \textbf{Without positional encoding (w/o PE)}: Evaluates the role of positional encoding. Removing PE results in a slight performance reduction, indicating that positional encoding provides continuity and contextual cues necessary for optimal scoring.
  
  \item \textbf{$M$ in LtE2E}: Evaluates the impact of $M$ in LtE2E. When $M=1$, performance suffers significantly due to insufficient pair diversity. Increasing $M$ to 50 and 200 substantially improves performance by exposing the model to more trials per batch and enriching training supervision. This demonstrates the effectiveness of LtE2E for end-to-end SV training.
  
\end{itemize}







%
\vspace{-2mm}
\begin{table}[!htbp]
\centering
\vspace{-2mm}
\caption{Results of ablation study.}
\vspace{-2mm}
\scalebox{0.73}{
\begin{tabular}{llcccccc}
\toprule
                               & \multicolumn{1}{}{}   & \multicolumn{2}{c}{Clean} & \multicolumn{2}{c}{Noisy} & \multicolumn{2}{c}{Mixing} \\
\cmidrule(r){3-4} \cmidrule(r){5-6} \cmidrule(r){7-8} & \multicolumn{1}{l}{\multirow{-2}{*}{Model}} & EER(\%)      & minDCF     & EER(\%)      & minDCF     & EER(\%)      & minDCF      \\
\midrule
& NS (M = 200)                                                      & 1.26        & 0.143      & 2.28        & 0.261      & 4.16        & 0.390       \\
\midrule
& ~~w/ shared encoder                                                     & 1.33        & 0.158      & 2.40        & 0.284      & 4.50        & 0.426       \\
& ~~w/ multiple layers                                                      & 1.60        & 0.162      & 2.64        & 0.277      & 4.51        & 0.433       \\
& ~~w/o PE                                                      & 1.30        & 0.150      & 2.35        & 0.273      & 4.35        & 0.408       \\
\midrule
& NS (M = 1)                                                        & 1.74        & 0.205      & 3.16        & 0.353      & 6.05        & 0.520       \\
& NS (M = 50)                                                       & 1.30        & 0.145      & 2.33        & 0.271      & 4.25        & 0.398       \\
\bottomrule
\end{tabular}}
\label{tab:ablation}
\vspace{-3mm}
\end{table}

\vspace{-1mm}
\section{Conclusion}
This paper presented Neural Scoring (NS), a novel end-to-end framework for speaker verification in complex acoustic environments, particularly in multi-speaker scenarios. Essentially, NS directly models the verification score as a function of both enrollment and test utterances, enabling a decoupled yet integrated and flexible learning process. To support effective optimization, we proposed a large-scale trial training strategy, which proved critical to superior model performance. 

Extensive experiments demonstrated that NS consistently achieved exceptional or near-optimal performance compared to all baselines and SOTA competitors.
Moreover, unlike other systems that suffer performance degradation from single- to multi-scenario training, NS benefited from diverse training data, showcasing strong generalization and robustness. Future work will focus on scaling up training to further amplify the advantages of the NS framework and assess its scalability in real-world deployment.

\newpage
\bibliographystyle{IEEEtran}
\bibliography{strings}

\begin{thebibliography}{10}
\providecommand{\url}[1]{#1}
\csname url@samestyle\endcsname
\providecommand{\newblock}{\relax}
\providecommand{\bibinfo}[2]{#2}
\providecommand{\BIBentrySTDinterwordspacing}{\spaceskip=0pt\relax}
\providecommand{\BIBentryALTinterwordstretchfactor}{4}
\providecommand{\BIBentryALTinterwordspacing}{\spaceskip=\fontdimen2\font plus
\BIBentryALTinterwordstretchfactor\fontdimen3\font minus \fontdimen4\font\relax}
\providecommand{\BIBforeignlanguage}[2]{{%
\expandafter\ifx\csname l@#1\endcsname\relax
\typeout{** WARNING: IEEEtran.bst: No hyphenation pattern has been}%
\typeout{** loaded for the language `#1'. Using the pattern for}%
\typeout{** the default language instead.}%
\else
\language=\csname l@#1\endcsname
\fi
#2}}
\providecommand{\BIBdecl}{\relax}
\BIBdecl

\bibitem{snyder2018x}
D.~Snyder, D.~Garcia-Romero, G.~Sell, D.~Povey, and S.~Khudanpur, ``X-vectors: Robust {DNN} embeddings for speaker recognition,'' in \emph{ICASSP 2018-2018 IEEE International Conference on Acoustics, Speech and Signal Processing (ICASSP)}.\hskip 1em plus 0.5em minus 0.4em\relax IEEE, 2018, pp. 5329--5333.

\bibitem{zeinali2019but}
H.~Zeinali, S.~Wang, A.~Silnova, P.~Mat{\v{e}}jka, and O.~Plchot, ``{BUT} system description to voxceleb speaker recognition challenge 2019,'' \emph{arXiv preprint arXiv:1910.12592}, 2019.

\bibitem{villalba2020state}
J.~Villalba, N.~Chen, D.~Snyder, D.~Garcia-Romero, A.~McCree, G.~Sell, J.~Borgstrom, L.~P. Garc{\'\i}a-Perera, F.~Richardson, R.~Dehak \emph{et~al.}, ``State-of-the-art speaker recognition with neural network embeddings in {NIST SRE18} and speakers in the wild evaluations,'' \emph{Computer Speech \& Language}, vol.~60, p. 101026, 2020.

\bibitem{li2022cn}
L.~Li, R.~Liu, J.~Kang, Y.~Fan, H.~Cui, Y.~Cai, R.~Vipperla, T.~F. Zheng, and D.~Wang, ``{CN-Celeb}: multi-genre speaker recognition,'' \emph{Speech Communication}, vol. 137, pp. 77--91, 2022.

\bibitem{chen2022build}
Z.~Chen, B.~Han, X.~Xiang, H.~Huang, B.~Liu, and Y.~Qian, ``Build a {SRE} challenge system: Lessons from {VoxSRC} 2022 and {CNSRC} 2022,'' in \emph{INTERSPEECH}, 2023, pp. 3202--3206.

\bibitem{qin2022simple}
X.~Qin, N.~Li, C.~Weng, D.~Su, and M.~Li, ``Simple attention module based speaker verification with iterative noisy label detection,'' in \emph{ICASSP 2022-2022 IEEE International Conference on Acoustics, Speech and Signal Processing (ICASSP)}.\hskip 1em plus 0.5em minus 0.4em\relax IEEE, 2022, pp. 6722--6726.

\bibitem{dehak2009support}
N.~Dehak, R.~Dehak, P.~Kenny, N.~Br{\"u}mmer, P.~Ouellet, and P.~Dumouchel, ``Support vector machines versus fast scoring in the low-dimensional total variability space for speaker verification,'' in \emph{INTERSPEECH}, 2009.

\bibitem{wang2023cam++}
H.~Wang, S.~Zheng, Y.~Chen, L.~Cheng, and Q.~Chen, ``{CAM++}: A fast and efficient network for speaker verification using context-aware masking,'' in \emph{INTERSPEECH}, 2023, pp. 5301--5305.

\bibitem{wang2023lightweight}
H.~Wang, X.~Lin, and J.~Zhang, ``A lightweight {CNN}-conformer model for automatic speaker verification,'' \emph{IEEE Signal Processing Letters}, vol.~31, pp. 56--60, 2023.

\bibitem{alam2023hybrid}
J.~Alam, W.~H. Kang, and A.~Fathan, ``Hybrid neural network with cross-and self-module attention pooling for text-independent speaker verification,'' in \emph{ICASSP 2023-2023 IEEE International Conference on Acoustics, Speech and Signal Processing (ICASSP)}.\hskip 1em plus 0.5em minus 0.4em\relax IEEE, 2023, pp. 1--5.

\bibitem{liu2024masv}
Y.~Liu, L.~Wan, Y.~Huang, M.~Sun, Y.~Shi, and F.~Metze, ``{MASV}: Speaker verification with global and local context mamba,'' \emph{arXiv preprint arXiv:2412.10989}, 2024.

\bibitem{heo2024next}
H.-J. Heo, U.-H. Shin, R.~Lee, Y.~Cheon, and H.-M. Park, ``{NeXt-TDNN}: Modernizing multi-scale temporal convolution backbone for speaker verification,'' in \emph{ICASSP 2024-2024 IEEE International Conference on Acoustics, Speech and Signal Processing (ICASSP)}.\hskip 1em plus 0.5em minus 0.4em\relax IEEE, 2024, pp. 11\,186--11\,190.

\bibitem{li2022adaptive}
R.~Li, S.~Fang, C.~Ma, and L.~Li, ``Adaptive rectangle loss for speaker verification.'' in \emph{INTERSPEECH}, 2022, pp. 301--305.

\bibitem{han2023exploring}
B.~Han, Z.~Chen, and Y.~Qian, ``Exploring binary classification loss for speaker verification,'' in \emph{ICASSP 2023-2023 IEEE International Conference on Acoustics, Speech and Signal Processing (ICASSP)}.\hskip 1em plus 0.5em minus 0.4em\relax IEEE, 2023, pp. 1--5.

\bibitem{liu2023awlloss}
Q.~Liu, X.~Zhang, X.~Liang, Y.~Qian, and S.~Yao, ``{AWLloss}: speaker verification based on the quality and difficulty of speech,'' \emph{IEEE Signal Processing Letters}, vol.~30, pp. 1337--1341, 2023.

\bibitem{cai2024leveraging}
D.~Cai and M.~Li, ``Leveraging {ASR} pretrained conformers for speaker verification through transfer learning and knowledge distillation,'' \emph{IEEE/ACM Transactions on Audio, Speech, and Language Processing}, 2024.

\bibitem{zhang2023adaptive}
L.~Zhang, Z.~Chen, and Y.~Qian, ``Adaptive large margin fine-tuning for robust speaker verification,'' in \emph{ICASSP 2023-2023 IEEE International Conference on Acoustics, Speech and Signal Processing (ICASSP)}.\hskip 1em plus 0.5em minus 0.4em\relax IEEE, 2023, pp. 1--5.

\bibitem{cai2023pretraining}
D.~Cai, W.~Wang, M.~Li, R.~Xia, and C.~Huang, ``Pretraining conformer with {ASR} for speaker verification,'' in \emph{ICASSP 2023-2023 IEEE International Conference on Acoustics, Speech and Signal Processing (ICASSP)}.\hskip 1em plus 0.5em minus 0.4em\relax IEEE, 2023, pp. 1--5.

\bibitem{choi2025trainable}
J.-H. Choi, J.-S. Seong, Y.-R. Jeoung, and J.-H. Chang, ``Trainable adaptive score normalization for automatic speaker verification,'' in \emph{ICASSP 2025-2025 IEEE International Conference on Acoustics, Speech and Signal Processing (ICASSP)}.\hskip 1em plus 0.5em minus 0.4em\relax IEEE, 2025, pp. 1--5.

\bibitem{snyder2019speaker}
D.~Snyder, D.~Garcia-Romero, G.~Sell, A.~McCree, D.~Povey, and S.~Khudanpur, ``Speaker recognition for multi-speaker conversations using x-vectors,'' in \emph{ICASSP 2019-2019 IEEE International Conference on Acoustics, Speech and Signal Processing (ICASSP)}.\hskip 1em plus 0.5em minus 0.4em\relax IEEE, 2019, pp. 5796--5800.

\bibitem{Rao2019TargetSE}
W.~Rao, C.~Xu, C.~E. Siong, and H.~Li, ``Target speaker extraction for multi-talker speaker verification,'' in \emph{INTERSPEECH}, 2019, pp. 1273--1277.

\bibitem{zhang2021towards}
C.~Zhang, M.~Yu, C.~Weng, and D.~Yu, ``Towards robust speaker verification with target speaker enhancement,'' in \emph{ICASSP 2021-2021 IEEE International Conference on Acoustics, Speech and Signal Processing (ICASSP)}.\hskip 1em plus 0.5em minus 0.4em\relax IEEE, 2021, pp. 6693--6697.

\bibitem{Shi2021}
Y.~Shi and T.~Hain, ``Supervised speaker embedding de-mixing in two-speaker environment,'' in \emph{2021 IEEE Spoken Language Technology Workshop (SLT)}.\hskip 1em plus 0.5em minus 0.4em\relax IEEE, 2021, pp. 758--765.

\bibitem{zhang2022enroll}
L.~Zhang, Z.~Chen, and Y.~Qian, ``Enroll-aware attentive statistics pooling for target speaker verification,'' in \emph{INTERSPEECH}, 2022, pp. 311--315.

\bibitem{heigold2016end}
G.~Heigold, I.~Moreno, S.~Bengio, and N.~Shazeer, ``End-to-end text-dependent speaker verification,'' in \emph{2016 IEEE international conference on acoustics, speech and signal processing (ICASSP)}.\hskip 1em plus 0.5em minus 0.4em\relax IEEE, 2016, pp. 5115--5119.

\bibitem{wen2022sphereface2}
Y.~Wen, W.~Liu, A.~Weller, B.~Raj, and R.~Singh, ``{SphereFace2}: Binary classification is all you need for deep face recognition,'' in \emph{The 10th International Conference on Learning Representations (ICLR 2022)}.\hskip 1em plus 0.5em minus 0.4em\relax OpenReview. net, 2022.

\bibitem{thienpondt2023margin}
J.~Thienpondt, N.~Madhu, and K.~Demuynck, ``Margin-mixup: A method for robust speaker verification in multi-speaker audio,'' in \emph{ICASSP 2023-2023 IEEE International Conference on Acoustics, Speech and Signal Processing (ICASSP)}.\hskip 1em plus 0.5em minus 0.4em\relax IEEE, 2023, pp. 1--5.

\bibitem{zhang2017mixup}
H.~Zhang, M.~Cisse, Y.~N. Dauphin, and D.~Lopez-Paz, ``Mixup: Beyond empirical risk minimization,'' in \emph{International Conference on Learning Representations}, 2018.

\bibitem{villalba2019state}
J.~Villalba, N.~Chen, D.~Snyder, D.~Garcia-Romero, A.~McCree, G.~Sell, J.~Borgstrom, F.~Richardson, S.~Shon, F.~Grondin \emph{et~al.}, ``State-of-the-art speaker recognition for telephone and video speech: The {JHU-MIT} submission for {NIST SRE18},'' in \emph{INTERSPEECH}, 2019, pp. 1488--1492.

\bibitem{aloradi2022speaker}
A.~Aloradi, W.~Mack, M.~Elminshawi, and E.~A. Habets, ``Speaker verification in multi-speaker environments using temporal feature fusion,'' in \emph{European Signal Processing Conference (EUSIPCO)}.\hskip 1em plus 0.5em minus 0.4em\relax IEEE, 2022, pp. 354--358.

\bibitem{van2008visualizing}
L.~Van~der Maaten and G.~Hinton, ``Visualizing data using {t-SNE}.'' \emph{Journal of machine learning research}, vol.~9, no.~11, 2008.

\bibitem{nagrani2017voxceleb}
A.~Nagrani, J.~S. Chung, and A.~Zisserman, ``Voxceleb: a large-scale speaker identification dataset,'' in \emph{INTERSPEECH}, 2017, pp. 2616--2620.

\bibitem{vaswani2017attention}
A.~Vaswani, N.~Shazeer, N.~Parmar, J.~Uszkoreit, L.~Jones, A.~N. Gomez, {\L}.~Kaiser, and I.~Polosukhin, ``Attention is all you need,'' \emph{Advances in neural information processing systems}, vol.~30, 2017.

\bibitem{chung2018voxceleb2}
J.~S. Chung, A.~Nagrani, and A.~Zisserman, ``Voxceleb2: Deep speaker recognition,'' in \emph{INTERSPEECH}, 2018, pp. 1086--1090.

\bibitem{snyder2015musan}
D.~Snyder, G.~Chen, and D.~Povey, ``Musan: A music, speech, and noise corpus,'' \emph{arXiv preprint arXiv:1510.08484}, 2015.

\bibitem{murtagh2012algorithms}
F.~Murtagh and P.~Contreras, ``Algorithms for hierarchical clustering: an overview,'' \emph{Wiley Interdisciplinary Reviews: Data Mining and Knowledge Discovery}, vol.~2, no.~1, pp. 86--97, 2012.

\end{thebibliography}

\end{document}